\documentclass[preprint,aps,amssymb,superscriptaddress,nofootinbib]{revtex4}
\usepackage{graphicx}
\usepackage{braket}
\newcommand{\del}{\partial}
\newcommand{\m}{\mathring}

\begin{document}

\title{Asymptotic Flatness and Quantum Geometry}

\author{Sandipan Sengupta}
\email{sandipan@rri.res.in}
\affiliation{Raman Research Institute\\
Bangalore-560080, INDIA.}

\begin{abstract}
 We construct a canonical quantization of the two dimensional theory of a parametrized scalar field on noncompact spatial slices. The kinematics is
built upon generalized charge-network states which are labelled by smooth
embedding spacetimes, unlike the standard basis states carrying only discrete labels. The resulting quantum geometry corresponds to a nondegenerate vacuum metric, which allows a consistent realization of the asymptotic conditions on the canonical fields. Although the quantum counterpart of the classical symmetry group of conformal isometries consists only of continuous global translations, Lorentz invariance can still be recovered in an effective sense. The quantum spacetime as characterised by a gauge invariant state is shown to be made up of discrete strips at the interior, and smooth at asymptotia. The analysis here is expected to be particularly relevant for a canonical quantization of asymptotically flat gravity based on generalized spin-network states labelled by smooth geometries. 
\end{abstract}


\maketitle

\section{Introduction}
It is generally anticipated that in any theory of quantum gravity, the notion of a smooth classical spacetime should give away to a discrete structure in the quantum regime. Such a description should be characterised by a fundamental length scale, which is expected to be the Planck length. In loop quantum gravity, which is one of the various approaches to a canonical quantization of gravity, the kinematical set up provides a natural realization of such discrete quantum geometries\cite{ashtekar1}. In this framework, the kinematical operators, namely the holonomies and fluxes, are manifestly diffeomorphism covariant. Remarkably, the geometrical operators depending on the flux, e.g. area, which carry the information associated with the spatial metric, exhibit discrete eigenspectra\cite{ashtekar,rovelli}. The kinematical ground state in this representation is the one which is annihilated by the flux operator, and hence corresponds to zero eigenvalue of area. Thus, the vacuum geometry is associated with a degenerate spatial metric. It should be emphasized that the spatial discreteness here is essentially a consequence of the choice of spin networks as the basic kinematical states, which carry discrete group-labels. 

However, within this quantization framework, it is difficult to conceive a way to recover smooth geometries\cite{nicolai}. The degenerate `vacuum' and the excitations on it lead to a structure which is far from a continuum. In particular, this description is not adequate for developing a quantization of asymptotically flat gravity, where the metric becomes flat as one approaches the spatial infinity. One can still expect that suitable coherent states or coarse-graining procedures can be constructed to capture the smoothness of the spatial geometry at asymptotia in an effective sense. However, such a success has not been realized as yet.

There have been earlier attempts to recover the notion of smooth geometries, keeping within the limits of standard LQG kinematics\cite{arnsdorf}. However, one can also explore a different approach based on a deformation of the standard representation, where the degenerate vacuum geometry is replaced by a smooth structure. Recently, a representation of this kind was introduced by Koslowski\cite{k} and further studied by Sahlmann\cite{s}. The essential construction here is based on spin-network states that carry a label corresponding to smooth spatial geometries, in addition to the discrete group-labels. The action of the flux operator on these is different than in the standard representation, and has an additional piece coming from the continuous background geometry:
\begin{eqnarray*}
\hat{E}_{s,f}\Psi~=~X_{s,f}\Psi~+~\m{E}_{s,f}I \Psi
\end{eqnarray*}
It was demonstrated that the formulation is diffeomorphism covariant\cite{s,mad,miguel}, keeping with the spirit of a background-independent quantization. To understand the implications of this representation for simpler systems, a loop (polymer) quantization of parametrized scalar field theory on a two dimensional Minkowskian cylinder (PFT) was constructed in \cite{sengupta}. Such an analysis was motivated by the fact that PFT serves as a solvable toy model of quantum gravity\cite{dirac,kuchar,l,laddha,laddha1}. The resulting quantum theory exhibits features which are remarkably different than the one based on a degenerate vacuum geometry as analysed by Laddha and Varadarajan\cite{laddha,laddha1}. Also, the eigenspectrum of the length operator was shown to be continuous, in contrast to the standard representation with a degenerate vacuum.

Here, armed with the insights gained from the analysis of PFT on a cylindrical spacetime in \cite{sengupta}, we proceed to set up a canonical quantization of PFT on noncompact spatial slices. This is done with a view to analyse the theory as a toy model for asymptotically flat gravity.

In the next section, we discuss the classical theory of asymptotically flat PFT in two dimensions. The boundary conditions to be imposed at infinity are spelled out in detail. In section III, we construct the quantum theory along the lines of ref.\cite{sengupta}, where the kinematics is based on charge-network states carrying a continuous label corresponding to smooth embeddings in addition to the original discrete labels. We also discuss the Dirac observables, and in particular, the conformal isometries. Section IV elucidates the basic features of the resulting quantum geometry. The final section contains a few relevant remarks.
\section{Classical theory}
\subsection{Action}
Relativistic field theories that are originally formulated on flat spacetime does not exhibit manifest spacetime covariance. However, such theories can be reformulated by the procedure known as `parametrization', as originally invoked by Dirac\cite{dirac}. In this form, the inertial spacetime coordinates become dynamical variables, and the general covariance becomes manifest- a feature which is shared by gravity theory. It was this perspective which led Kuchar to analyse the theory of a parametrized scalar field on a two dimensional Minkowskian cylinder as a toy model for four dimensional gravity theory\cite{kuchar}.

Free scalar field theory in $1+1$ dimensional Minkowskian spacetime is described by the following action:
\begin{eqnarray}\label{s}
S[\phi(X)]~=~\int d^2 X ~\eta^{AB}\del_A \phi(X)\del_B \phi(X)
\end{eqnarray}
where, $\phi(X^A)$ are the (dynamical) scalar fields; $X^A,~A=0,1,$ are the (non-dynamical) inertial coordinates corresponding to the flat spacetime; $\eta_{AB}\equiv [-1,1]$ is the flat metric. A manifestly covariant formulation of the same theory can be obtained by treating the inertial coordinates $X^A(x)$ as additional dynamical fields besides $\phi(x)$. These fields are now parametrized by the arbitrary coordinates $x^\alpha$. The corresponding action becomes:
\begin{eqnarray}\label{s-pft} 
S_{PFT}[X^A(x),\phi(x)]~=~\int d^2 x ~g(X)^{\frac{1}{2}}g^{\alpha\beta}(X)\del_\alpha \phi(x)\del_\beta \phi(x)
\end{eqnarray}
where, $g(X)$ is the determinant of the metric $g_{\alpha\beta}(X)$, given by:
\begin{eqnarray*}
g_{\alpha\beta}(X)~=~\eta_{AB} \frac{\del X^A(x)}{\del x^\alpha}\frac{\del X^B(x)}{\del x^\beta}
\end{eqnarray*}
The action (\ref{s-pft}) is covariant under two dimensional spacetime diffeomorphisms. This is a reflection of the fact that the dynamical content of the theory does not depend on the choice of coordinates $x^\alpha$. Evidently, one can freeze the two functions' worth of gauge freedom by choosing $X^A$ as $x^\alpha$ and recover (\ref{s}) from (\ref{s-pft}). Thus, the two formulations have the same number of physical degrees of freedom.

\subsection{Hamiltonian formulation}
Here we take the spatial coordinate $x_1=x$ is to be noncompact, in order to construct an analogue of the case of four dimensional gravity on noncompact Cauchy slices. We assume that each $t=const$ slice corresponds to an one dimensional spacelike surface $X^A(t,x)$, defining an arbitrary foliation of the spacetime. In the Hamiltonian theory based on (\ref{s-pft}), the basic canonical pairs in the embedding and matter sectors are  $(X^A(x),~\Pi_A(x))$ and $(\phi(x),~\pi(x))$. It is convenient to redefine these variables and work with two mutually exclusive sectors of left and right moving fields:
\begin{eqnarray*}
&&X^\pm(x)~=~T(x)\pm X(x),~~\Pi_\pm(x)~=~\frac{1}{2}\left[\Pi_0(x) \pm \Pi_1(x)\right]\\
&& Y^\pm~=~\pi(x)\pm \phi'(x)
\end{eqnarray*}
The basic Poisson-brackets between these are listed below:
\begin{eqnarray*}
&&\left\{ X^\pm(x),\Pi_\pm(y)\right\}~=~\delta(x,y),~~\left\{ X^\mp(x),\Pi_\pm(y)\right\}~=~0,\\
&&\left\{ Y^\pm(x),Y^\pm(y)\right\}~=~\pm [\del_x\delta(x,y)-\del_y\delta(x,y)],~~\left\{ Y^\mp(x),Y^\pm(y)\right\}~=~0
\end{eqnarray*}
The gauge freedom underlying the action (\ref{s-pft}) leads to a pair of first-class constraints:
\begin{eqnarray*}
H_\pm~=~\Pi_\pm(x) X^{\pm'}(x)\pm\frac{1}{4}Y^\pm(x)^2
\end{eqnarray*}
These obey the following algebra:
\begin{eqnarray*}
\left\{\int dx N^\pm(x) H_\pm(x),\int dx M^{\pm}(y)H_{\pm}(y)\right\}~&=&~\int dx [N^\pm (x)M^{\pm '}(x)-M^\pm(x)N^{\pm '}(x)] H_\pm(x)\\
\left\{\int dx N^\mp(x) H_\mp(x),\int dx M^\pm(y)H_\pm(y)\right\}~&=&~0
\end{eqnarray*}
where, $N^\pm$ and $M^\pm$ are Lagrange multipliers.
Their action can be interpreted as two independent spatial diffeomorphisms in the left and right moving sectors\cite{laddha,laddha1}.
Thus, the action of finite diffeomorphisms on the fields can be characterised by a pair  $(\xi^+,\xi^-)$:
\begin{eqnarray*}
\left(\alpha_{\xi^\pm}X^\pm\right)(x)~&=&~X^\pm(\xi^\pm(x))\\
\left(\alpha_{\xi^\mp}X^\pm\right)(x)~&=&~X^\pm(x)\\
\left(\alpha_{\xi^\pm}F^\pm\right)(x)~&=&~F^\pm(\xi^\pm(x))\\
\left(\alpha_{\xi^\mp}F^\pm\right)(x)~&=&~F^\pm(x)~.
\end{eqnarray*}
%
\subsection{Asymptotic behaviour of fields}
Since the spatial slice is noncompact, one needs to impose suitable conditions on the fields at the spatial infinity so that the symplectic form, canonical constraints and observables do not lead to divergent expressions or to a violation of functional differentiability\cite{regge}. 
Demanding that the spatial slice asymptotes to the flat slice, we obtain:
\begin{eqnarray}\label{Xfalloff}
&& X^{\pm}(x)~=~ e^{\pm \lambda_R} x~+~\mu_R^{\pm}~+~\frac{\nu^{\pm}_R}{x}~+~O\left(\frac{1}{x^2}\right) \mathrm{~~~as~x\rightarrow +\infty~~,} \nonumber \\
&& X^{\pm}(x)~=~ e^{\pm \lambda_L} x ~+~\mu_L^{\pm}~+~\frac{\nu^{\pm}_L}{x}~+~O\left(\frac{1}{x^2}\right)\mathrm{~~~as~x\rightarrow -\infty~~.}
\end{eqnarray}
Here $\lambda_{R,L},~\mu_{R,L}^{\pm},~\nu_{R,L}^{\pm}$ are constants and the remainder has the property that $\lim_{x\rightarrow\infty} x~O\left(\frac{1}{x^2}\right)\rightarrow 0$.
Note that this leads to the following fall-off of the spatial metric $g_{xx}=X^{+'}(x)X^{-'}(x)$:
\begin{eqnarray*}
g_{xx}~-~\eta_{xx}~=~O\left(\frac{1}{x^2}\right)
\end{eqnarray*}
The action of the cotangent vectors $\Pi_{\pm}(x)$ on the tangent vectors $\delta X^{\pm}(x)$ is well defined if $\int_{\Sigma_t}\Pi_{\pm} \delta X^{\pm}$ is finite, where $\Sigma_t$ denotes the spatial slice. Here $\delta X^{\pm}(x)$ is defined as the infinitesimal variation of the solutions $X^{\pm}(x)$:
\begin{eqnarray*}
\delta X^{\pm}(x)~&=&~\pm e^{\pm \lambda_R}x \delta\lambda_{R} ~+~\delta\mu_R^{\pm}~+~O\left(\frac{1}{r}\right)\mathrm{~~~as~x\rightarrow +\infty~~,} \nonumber\\
\delta X^{\pm}(x)~&=&~\pm e^{\pm \lambda_L}x \delta\lambda_{L} ~+~\delta\mu_L^{\pm}~+~O\left(\frac{1}{r}\right)\mathrm{~~~as~x\rightarrow -\infty~~.}
\end{eqnarray*}
 Thus, we obtain:
\begin{eqnarray}
\Pi(x)\sim \frac{1}{x^3} \mathrm{~~~as~x\rightarrow \pm\infty~.}
\end{eqnarray}
The embedding parts of the smeared constraints are functionally differentiable and preserve the above falloffs of the fields if we make the following choice:
\begin{eqnarray}\label{Nfall}
N_{\pm}(x) ~&=&~\pm \alpha_R x~+~\beta_R^{\pm}~+~O\left(\frac{1}{x}\right)\mathrm{~~~as~x\rightarrow +\infty~~,}\nonumber\\ 
 N_{\pm}(x) ~&=&~\pm \alpha_L x~+~\beta_L^{\pm}~+~O\left(\frac{1}{x}\right)\mathrm{~~~as~x\rightarrow -\infty~~.}
\end{eqnarray}
Here, an important remark is in order. In (\ref{Nfall}), the first two terms in the expansion correspond to asymptotic boosts and translations. Although the corresponding vector fields do not die away at infinity, the smeared constraint functionals $\int N^{\pm}H_{\pm}$ are still well-defined. Thus, these represent `proper' gauge transformations\cite{marolf}. This is very different from the case of asymptotically flat gravity, where the translations, boosts and rotations at the spatial infinity do not leave the constraint functional well defined and functionally differentiable. In that case, one needs to supplement the total Hamiltonian with an appropriate boundary term, and the resulting functional is not a constraint anymore since it does not vanish on-shell. Thus, it cannot be interpreted as the generator of gauge transformations\cite{regge}. For example, the asymptotic translations in this case generate `proper' time evolutions at spatial infinity, unlike the ones in PFT which do not generate any physical evolution of the phase space data. 

Finally, using a similar analysis for the matter part of the smeared constraints, we obtain:
\begin{eqnarray}\label{matterfall} 
\phi(x) &\sim & \frac{1}{x}~,\nonumber\\
\pi(x) &\sim & \frac{1}{x^2}~,\nonumber\\
Y^{\pm}(x)&\sim & \frac{1}{x^2} \mathrm{~~~as~x\rightarrow \pm\infty~~.}
\end{eqnarray}
 The conditions (\ref{matterfall}) also ensure that the scalar field symplectic form is well defined. 
\subsection{Matter dependent Dirac observables}
The space-integral of a scalar density of weight one, which can be written in terms of the phase-space variables separately in the `$+$' or `$-$' sector, is a Dirac observable, and hence commutes with the constraints. An example is\cite{laddha,laddha1}:
\begin{eqnarray}\label{dirac}
O_f^{\pm}~=~\int dx~ Y^{\pm}(x) f^{\pm}(X^\pm)(x)
\end{eqnarray}
where $f^{\pm}(X^\pm)$ are real functions of $X^\pm$. $O_f^{\pm}$ is well-defined if the following holds:
\begin{eqnarray*}
f^{\pm}(X^{\pm})(x)~&=&~ c_R^{\pm}~+~O\left(\frac{1}{x}\right)\mathrm{~~~~~as~x\rightarrow +\infty,}\\
f^{\pm}(X^{\pm})(x)~&=&~c_L^{\pm}~+~O\left(\frac{1}{x}\right)\mathrm{~~~~~as~x\rightarrow -\infty}
\end{eqnarray*}
where, $c^{\pm}_{R,L}$ are constants.
%

\subsection{Conformal isometries}
There is another class of observables, which, unlike the above, depend only on the embedding sector. These are the generators of local Weyl rescalings of the metric $g_{\alpha\beta}$:
\begin{eqnarray*}
O_c^{\pm}~=~\int dx~ \Pi_{\pm}(x) U^{\pm}(X^{\pm})(x)
\end{eqnarray*}
where, $U^{\pm}(X^{\pm})$ is any function of $X^{\pm}$ such that
\begin{eqnarray*}
U^{\pm}(X^{\pm})(x)~&=&~\pm\gamma_R  x~+~\zeta_R^{\pm}~+~O\left(\frac{1}{x}\right)~~~\mathrm{~~~~~as~x\rightarrow +\infty,}\\
U^{\pm}(X^{\pm})(x)~&=&~\pm\gamma_L  x~+~\zeta_L^{\pm}~+~O\left(\frac{1}{x}\right)~~~\mathrm{~~~~~as~x\rightarrow +\infty}~.
\end{eqnarray*}
Here $ \gamma_{R,L}^{\pm},~\zeta_{R,L}$ are constants.

 The special cases $U^{\pm}(X^{\pm})=\zeta^{\pm}$ and $U^{\pm}(X^{\pm})=e^{\pm \lambda}X^{\pm}$ for constant parameters $\zeta^{\pm},~\lambda$ are of particular interest. These generate global translations and boosts respectively, which completely characterise the global Poincare group acting on $X^{\pm}$.

The group of conformal isometries also include asymptotic translations and boosts. These correspond to the following conditions, respectively:
\begin{eqnarray*}
&&U^{\pm}(X^{\pm})=c_R^{\pm}\mathrm{~~~~as~x\rightarrow +\infty,~~}U^{\pm}(X^{\pm})=c_L^{\pm}~\mathrm{~~~~as~x\rightarrow -\infty};\\
&&U^{\pm}(X^{\pm})=e^{\pm c_R^{'}}X^{\pm}\mathrm{~~~as~x\rightarrow +\infty,~~} U^{\pm}(X^{\pm})=e^{\pm c_L^{'}} X^{\pm}\mathrm{~~as~x\rightarrow -\infty}~.
\end{eqnarray*} 
Note that these asymptotic conditions do not imply any restriction on the possible form of $U^{\pm}(X^{\pm})$ at the interior.

The infinitesimal isometry generators satisfy the following algebra:
\begin{eqnarray*}
\left\{\int dx ~\Pi_{\pm}(x) U^{\pm}(X^{\pm})(x),\int dy~ \Pi_{\pm}(y) V^{\pm}(X^{\pm})(y)\right\}=\int dx~ \Pi_{\pm}(x) (U^{\pm}V^{\pm'}-V^\pm U^{\pm'})(x)
\end{eqnarray*}
The finite action of these can be represented  through a pair of functions $\phi_c^{\pm}$ which act only on the embedding sector:
\begin{eqnarray*}
&&\alpha_{\phi^{\pm}_c}  X^{\pm}(x)=\phi^{\pm}_c(X^{\pm}(x)),~~\alpha_{\phi^{\mp}_c}  X^{\pm}(x)=X^{\pm}(x),\\
&&\alpha_{\phi^{\pm}_c}  Y^{\pm}(x)= Y^{\pm}(x),~~\alpha_{\phi^{\mp}_c}  Y^{\pm}(x)= Y^{\pm}(x)
\end{eqnarray*}
The functions $\phi_c^\pm(X^\pm)$ are invertible, connected to identity, and monotonically increasing:
\begin{eqnarray*}
\frac{d\phi_{c}^\pm}{dX^\pm}>0
\end{eqnarray*}
This ensures that if the original spatial metric $g_{xx}$ is nondegenerate, then so is the final one obtained from the map $\phi_c^\pm(X^\pm)$.
 %
%
\section{Quantum theory}
Based on the Hamiltonian theory, we now proceed to develop a canonical quantization of PFT. We follow the kinematical construction as adopted in ref.\cite{sengupta}.
The kinematical states are the charge-network states, defined separately in the embedding and matter sectors. 
\subsection{Embedding sector}
The embedding charge network $\ket{s_e^{\pm},\m{X^{\pm}}}:=\ket{\gamma_e^{\pm},\overrightarrow{k^{\pm}},\m{X}^{\pm}}$
is defined on a pair of graphs $\gamma_e^{\pm}$. Each of these graphs are composed of non-overlapping (except at the vertices) edges. The $i$-th edge $e_i^{\pm}$ has an embedding charge $k_i^{\pm}=\alpha n^{\pm},~n^{\pm}\in Z$. The fixed real number $\alpha$ is a free parameter  in the theory and is an analogue of the Barbero-Immirzi parameter in LQG. The embedding charge network is labelled by $\gamma_e^{\pm}$, a set of embedding charges $\overrightarrow{k^{\pm}}=(k_1^{\pm},..,k_{N^{\pm}}^{\pm})$ and smooth embedding charges $\m{X}^{\pm}(x)$. To emphasize, $\m{X}^{\pm}(x)$ has no analogue in the standard polymer representation based on a degenerate vacuum metric, where states are labelled only by discrete charges.

To capture the notion of the smooth asymptotic geometry through the kinematical states, we assume that the discrete embedding charges attain constant values outside the compact region $x\in[-L_e^{\pm},L_e^{\pm}]$ which is spanned by a finite number of edges of the embedding graph $\gamma_e^{\pm}$: 
\begin{eqnarray*}
k_i^{\pm}&=& k_L^{\pm} \mathrm{~~~for~ i<1},\\
 k_i^{\pm}&=& k_R^{\pm} \mathrm{~~~for~ i>N^{\pm}}.
\end{eqnarray*}
Notice that although the charge network extends out to infinity, the vertices do not, since they always lie within a compact region.

The inner product between any two charge-nets is given by:
\begin{eqnarray*}
\braket{s^{'\pm}_e,\m{X}^{'\pm}|s^{\pm}_e,\m{X}^{\pm}}=\delta_{s^{'\pm}_e,s_e^{\pm}} ~\delta_{\m{X}^{'\pm},\m{X}^{\pm}}
 \end{eqnarray*}
 The basic operators are the embedding fields $\hat{X}^{\pm}(x)$ and the holonomies $\hat{h}_{\gamma_e^{\pm}}=e^{i\sum_{i=1}^{N^{\pm}} k^{\pm}_{i} \int_{e_i\in \gamma_e} \Pi_{\pm}}$ defined on the graph $\gamma_e^{\pm}$. The motivation for such a choice of basic operators stems from LQG, where the kinematics is based on the holonomy and flux operators which are manifestly diffeomorphism covariant. 
Their action on the charge-network states are as follows:
\begin{eqnarray}
\hat{X}^{\pm}(x)\ket{\gamma_e^{\pm},\overrightarrow{k^{\pm}},\m{X}^{\pm}}&=&(k^{\pm}_{x,s_e^{\pm}}+\m{X}^{\pm}(x))\ket{\gamma_e^{\pm},\overrightarrow{k^{\pm}},\m{X}^{\pm}}\nonumber \\
 \hat{h}_{\gamma_{e}^{'\pm}}\ket{\gamma_e^{\pm},\overrightarrow{k^{\pm}},\m{X}^{\pm}}&=&\ket{\gamma_e^{'\pm} \circ\gamma_e^{\pm},\overrightarrow{k^{'\pm}}+\overrightarrow{k^\pm},\m{X}^{\pm}}
\end{eqnarray} 
where \begin{eqnarray*}
k^{\pm}_{x,s_e^{\pm}}&=&k^{\pm}_m \mathrm{~~~~if ~x \in Interior(e^{\pm}_m)}~,\\
&=&\frac{1}{2}(k^{\pm}_m+k^{\pm}_{m+1}) \mathrm{~~~~if~x \in e^{\pm}_m \cap e^{\pm}_{m+1}}~,\\
&=& k_L\mathrm{~~~if~x<-L_e^{\pm}}~, \\
&=& k_R\mathrm{~~~if~x>L_e^{\pm}}
\end{eqnarray*}
Here we have defined $\gamma_e^{'\pm}\circ\gamma_e^{\pm}$ as a graph finer than both $\gamma_e^{'\pm}$ and $\gamma_e^{\pm}$. The edges of $\gamma_e^{'\pm}\circ\gamma_e^{\pm}$ are characterised by the following distribution of embedding charges: \\
(a) In regions (of $\gamma^{'\pm}_{e}\circ  \gamma^{\pm}_{e}$) where the edges of $\gamma^{'\pm}_{e}$ and $\gamma^{\pm}_{e}$ do not overlap, the charges are the same as the original ones;\\
(b) In regions where the edges overlap, the charge is the sum of the original charges.\\
Also, we identify any two sets of embedding data $[k_{x,s_e^{\pm}}^{\pm},\m{X}^{\pm}(x)]$ and $[k_{x,s_e^{\pm}}^{\pm}+n^{\pm},\m{X}^{\pm}(x)-n^{\pm}]$ to be the same, where $n^{\pm}$ are integers.

\subsection{Matter sector:}
Motivated by the construction of operators in the embedding sector, we choose, instead of the matter fields $Y^{\pm}(x)$, the corresponding holonomies  $\hat{h}_{\gamma_m^{\pm}}$ as the basic operators\cite{laddha,laddha1}. These are defined on the graphs $\gamma_m^{\pm}$ as:
\begin{eqnarray}\label{holonomy-m}
\hat{h}_{\gamma_m^{\pm}}=e^{i\sum_{i=1}^{N^{\pm}} l^{\pm}_{i} \int_{e_i \in \gamma_m} Y^{\pm}}\end{eqnarray}
Each edge $e_i^{\pm}$ carries a matter charge $l^{\pm}_i$. These charges are of the form $l^{\pm}_i=\beta n^{\pm}$, where $\beta$ is a real constant of dimension $h^{-\frac{1}{2}}$ and $n$ are integers. The charge network states for the matter sector $\ket{s_m^{\pm}}:=\ket{\gamma_m^{\pm},\overrightarrow{l^{\pm}}}$ are labelled by $\gamma_m^{\pm}$ and the set $\overrightarrow{l^{\pm}}=(l_1^{\pm},..,l^{\pm}_N)$. As in the embedding sector, we assume that the matter charges take discrete values only within a compact region $x\in[-L_m^{\pm},L_m^{\pm}]$ which is covered by a finite number of non-overlapping edges of the matter graph $\gamma_m^{\pm}$ and are constant outside: 
\begin{eqnarray*}
&&l_j^{\pm}=l_L^{\pm} \mathrm{~~~for~ j<1} \\
&&l_j^{\pm}=l_R^{\pm}\mathrm{~~~for~j>M}.
\end{eqnarray*}
On these charge-network states, the matter-holonomy operators act as:
 \begin{eqnarray}\label{hm}\hat{h}_{\gamma_m^{'\pm}} \ket{\gamma_m^{\pm},\overrightarrow{l^{\pm}}}=e^{\frac{i}{2}\theta(l^{\pm},
 l^{'\pm})}\ket{\gamma_m^{'\pm}\circ\gamma_m^{\pm},~\overrightarrow{l^{'\pm}}+\overrightarrow{l^{\pm}}}
\end{eqnarray} 
Here the graph $\gamma_m^{'\pm}\circ\gamma_m^{\pm}$ is finer than both $\gamma_m^{'\pm}$ and $\gamma_m^{\pm}$, and is defined exactly as in the embedding sector. Note that the real number (phase) $\theta(l^{\pm},l^{'\pm})$, which arises due to the noncommutativity of $Y^{\pm}(x)$'s, is bilinear in the matter charges $l^{\pm},l^{'\pm}$. For example, in the simplest case where the state $\ket{\gamma_m^{\pm},\overrightarrow{l^{\pm}}}$ and the holonomy $\hat{h}_{\gamma_m^{'\pm}}$ in eq.(\ref{hm}) are based on the same graph $\gamma_m^{\pm}$, this becomes:
\begin{eqnarray*} 
\theta(l^{\pm},l^{'\pm})=\sum_{i=1}^{N^{\pm}-1} \left(l^{\pm}_{i} l^{'\pm}_{i+1}-
l^{'\pm}_{i} l^{\pm}_{i+1}\right)~~.
\end{eqnarray*}
\subsection{Nondegeneracy condition}
The spatial metric metric $g_{xx}=X^{+'}(x)X^{-'}(x)$ is assumed to be nondegenerate in the classical theory, implying $\pm X^{\pm'}(x)>0$. This leads to the following conditions in the quantum theory:

\begin{eqnarray}\label{nondeg}
&&(a)~\pm(k_{m^{\pm}+1}^{\pm}-k_{m^{\pm}}^{\pm}) \geqslant 0 \mathrm{~for ~1\leqslant m^{\pm}\leqslant (N^{\pm} -1)~~~and} \nonumber\\
&&(b)~\pm k_{1}^{\pm}\geqslant k_{L}^{\pm} ,~~\pm k_N^{\pm}\leqslant k_{R}^{\pm}.
\end{eqnarray} 
This implies that the nondegeneracy of the spatial metric $g_{xx}$ at all points is preserved in the quantum theory.

\subsection{Unitary action of gauge transformations}
The action of the finite gauge transformations $(\xi^+,\xi^-)$ can be represented by a pair of unitary operators $(\hat{U}_{\xi^+},\hat{U}_{\xi^-})$:
\begin{eqnarray}
 \hat{U}_{\xi^{\pm}} \ket{s_e^{\pm},\m{X}^{\pm}}\otimes \ket{s_m^{\pm}}~=~\ket{\xi^{\pm}(s_e^{\pm}),\xi_*^{\pm} \m{X}^{\pm}}\otimes \ket{\xi^{\pm}(s_m^{\pm})}
 \end{eqnarray}
 where $\xi^{\pm}(s^{\pm})$ is the image of $s^{\pm}$ under the action of $\hat{U}_{\xi^{\pm}}$ and $\xi^{\pm}_* \m{X}^{\pm}$ denotes the push-forward of $\m{X}^{\pm}$. The embedding charges $k^{\pm}_{x,\xi^{\pm}(s_e^{\pm})}$ in the new charge-network are related to the original ones as:
 \begin{eqnarray}
 k^{\pm}_{x,\xi^{\pm}(s_e^{\pm})}~=~k^{\pm}_{\xi^{{\pm}^{-1}}(x),s_e^{\pm}}
 \end{eqnarray}
 It is straightforward to see that the above action is a representation and induces the correct transformation of the basic operators. The relevant details are very similar to the case of PFT in a cylindrical spacetime as analysed in ref.\cite{sengupta}, and hence would be omitted here.
\subsection{Matter dependent Dirac observables}
Although the classical Dirac observables $O_f^{\pm}$ in (\ref{dirac}) cannot be represented in terms of the basic operators, their exponentials  can be. They have an unitary action on the kinematical states:
\begin{eqnarray}
 &&\widehat{e^{i\int f^{\pm}(X^{\pm}(x)) Y^{\pm}(x)}}\ket{\gamma_e^{\pm},\overrightarrow{k^{\pm}},\m{X}^{\pm}}\otimes\ket{\gamma_m^{\pm},\overrightarrow{l^{\pm}}}\nonumber\\
 &=& \widehat{e^{i\int f^{\pm}(k_{x,s_e^{\pm}}^{\pm}+\m{X}^{\pm}(x)) Y^{\pm}(x)}}\ket{\gamma_e^{\pm},\overrightarrow{k^{\pm}},\m{X}^{\pm}}\otimes\ket{\gamma_m^{\pm},\overrightarrow{l^{\pm}}}\nonumber\\
 &=& \hat{h}_{\gamma^{\pm}_{m_{f}}}~\ket{\gamma_e^{\pm},\overrightarrow{k^{\pm}},\m{X}^{\pm}} \otimes \ket{\gamma_m^{\pm},\overrightarrow{l^{\pm}}}\nonumber\\
 &=& e^{\frac{i}{2}\theta(l^{\pm},f^{\pm})}\ket{\gamma_e^{\pm},\overrightarrow{k^{\pm}},\m{X}^{\pm}}\nonumber \otimes \ket{\gamma_{m_{f}}^{\pm} \circ \gamma_m^{\pm},~\overrightarrow{l^{\pm}}+\overrightarrow{f^{\pm}}(k^{\pm}+\m{X}^{\pm})}
\end{eqnarray}
where, the graph $\gamma_{m_{f}}^{\pm} o  \gamma_{m}^{\pm}$ is finer than both $\gamma_{m_{f}}^{\pm}$ and $\gamma_{m}^{\pm}$ and we define the matter holonomy operator $\hat{h}_{\gamma^{\pm}_{m_{f}}}$ as:
\begin{eqnarray} \label{h-f}
\hat{h}_{\gamma^{\pm}_{m_{f}}}=\widehat{e^{i\int f^{\pm}(k_{x,s}^{\pm}+\m{X}^{\pm} (x)) Y^{\pm}(x)}}
\end{eqnarray} 
In order to have a well-defined action on the matter sector, the function $f^{\pm}(X^{\pm})$ must be piecewise constant at the interior. Outside this compact region, it should take constant values, which can be independent for the `$+$' and `$-$' sectors as well as at the left and right infinity. In general, the mapping $f^{\pm}$ should satisfy the following condition: \begin{eqnarray*} f^{\pm}(\alpha n^{\pm}+\m{X}^{\pm})=\beta n^{'\pm}
\end{eqnarray*} 
with $n^{\pm},n^{'\pm} \in Z$.
\subsection{Conformal isometries}
To represent the action of the conformal isometries $\phi_c:=(\phi^{+}_c,\phi^{-}_c)$, we can choose a pair of unitary operators $\hat{V}_{\phi_c^{\pm}}$ which has the following action:
\begin{eqnarray}\label{v-phi}
 \hat{V}_{\phi_{c}^{\pm}}\ket{\gamma_e^{\pm},\overrightarrow{k},\m{X}^{\pm}}\otimes \ket{\gamma_m^{\pm},\overrightarrow{l}}~=~ \ket{\gamma^{\pm}_{e},\overrightarrow{\bar{k}},\m{\bar{X}}^{\pm}}\otimes \ket{\gamma_m^{\pm},\overrightarrow{l}}
\end{eqnarray}
where, at any point $x$ in the graph $\gamma_e$, the embedding charges $\overrightarrow{(\bar{k}}^{\pm},\m{\bar{X}}^{\pm})$ in the new charge-net $\ket{\gamma^{\pm}_{e},\overrightarrow{\bar{k}},\m{\bar{X}}^{\pm}}$ are related to the original ones  through the mapping $\phi_c^{\pm}(X^{\pm})$ as:
\begin{eqnarray}\label{conf}
 \bar{k}^{\pm}+~\m{\bar{X}}^{\pm}=\phi_c^{\pm^{-1}}(k^{\pm}+\m{X}^{\pm})
\end{eqnarray}

From the above, it follows that
\begin{eqnarray*}
\hat{V}_{\phi_c^{\pm}} \hat{X}^{\pm}(x)\hat{V}^{\dagger}_{\phi_c^{\pm}}\ket{\gamma_e^{\pm},\overrightarrow{k},\m{X}^{\pm}}\otimes \ket{\gamma_m^{\pm},\overrightarrow{l}}
=\phi^{\pm}_c(\hat{X}(x))\ket{\gamma_e^{\pm},\overrightarrow{k},\m{X}^{\pm}}\otimes \ket{\gamma_m^{\pm},\overrightarrow{l}}
\end{eqnarray*}
which mimics the classical transformation law. Next, we analyse to what extent the classical group of conformal isometries can be represented through the invertible and monotonic maps $\phi_c^{\pm}(X^{\pm})$ in the quantum theory.

Let us consider $\phi_c(X)$ to be a nonlinear map given by $\phi_c(X)=X^n$, where $n$ is any integer. We denote the initial and final differences between the total embedding charges across any vertex $v$ by $\Delta=\alpha n$ and $\Delta'=\alpha n'$, where $n,n'\in Z$ and $\alpha$ is a fixed real number as defined earlier. Also, just for convenience, we choose $\alpha=1$. Now, assuming that the discrete embedding charges are $k$ and $k'$ respectively across $v$, and using the continuity of $\m{X}$ at $v$, we obtain:
 \begin{eqnarray*}
 \Delta'&=& \lim_{\epsilon\rightarrow 0}\left[\left(k'+\m{X}(v+\epsilon)\right)^n ~-~\left(k+\m{X}(v-\epsilon)\right)^n \right]\\
 &=& (k'^n-k^n)+c_1(k'^{n-1}-k^{n-1})\m{X}(v)+c_2(k'^{n-2}-k^{n-2})\m{X}(v)^2+...+c_{n-1}(k'-k)\m{X}(v)^{n-1}
 \end{eqnarray*}
 where, the $c_n$-s are the binomial coefficients, which are integers.
 Now notice that only the first term is an integer, whereas for any general $\m{X}(v)\in R$, the rest are all real-valued, except for the special case $\m{X}(v)^{p}=\frac{m_p}{c_i(k'^{n-p}-k^{n-p})},~m_p \in Z$ for all $p=1,..,n-1$. However, such a particular solution is not stable under any arbitrarily small (real-valued) perturbation of the smooth function $\m{X}$,  since that leads to a noninteger $\Delta'$. Thus, although the jump across $v$ was integer-valued to begin with, under a general nonlinear mapping $\phi_c$ and for any general $\m{X}$, the jump would not be integer-valued anymore. This argument also applies to any linear combination of functions of the form $\phi_c(X)=X^n$.   
Thus, eq.(\ref{conf}) can be satisfied only for linear maps $\phi_c^{\pm}(X^{\pm})$. Global translations and boosts are the only examples of such maps. We consider these separately below:
\vspace{.2cm}\\
{\bf Translations:}
\vspace{.1cm}\\
These can be implemented on the states in a  straightforward manner. Their action can be characterised by any two real numbers $r^{\pm}$:
\begin{eqnarray*}
 \phi_c^{\pm}(k^{\pm}_{x,s_e}+\m{X}^{\pm} (x))~=~k^{\pm}_{x,s_e}+\m{X}^{\pm} (x)+r^{\pm}
 \end{eqnarray*}
 Since $r^{\pm}$ are arbitrary, the group of translations is continuous.
 \vspace{.2cm}\\
{\bf Boosts:}
\vspace{.1cm}\\
Under finite global boosts parametrised by a constant number $\lambda$, the states should transform as: 
\begin{eqnarray*}
\hat{V}_{\lambda}\ket{\gamma_e^{\pm},\overrightarrow{k^{\pm}},\m{X}^{\pm}}=\ket{\gamma_e^{\pm},e^{\pm \lambda}\overrightarrow{k^{\pm}},e^{\pm \lambda}\m{X}^{\pm}}
\end{eqnarray*}
As earlier, we consider the difference $\Delta^{\pm}$ between the total embedding charge $[k_{x^{\pm},s_e^{\pm}}+\m{X}^{\pm}(x)]$ across any vertex $v$. Since the continuous charge $\m{X}^{\pm}$ does not contribute to this difference, we obtain $\Delta^{\pm}=\Delta k^{\pm}|_{v}=\alpha n^{\pm}$, where $n^{\pm}$ are integers. Now, after applying the boost, the new value of the spacing becomes $\Delta^{\pm}=\alpha m^{\pm}$, where $m^{\pm}$ are new integers, different from $n^{\pm}$ in general. Thus, we obtain:
\begin{eqnarray*}
e^{\pm \lambda}=\frac{m^{\pm}}{n^{\pm}}
\end{eqnarray*}
Thus, the boost parameter $e^{\pm \lambda}$ have to be a rational number, which must be the same for all vertices. Such a boost is state dependent, and cannot be applied to any arbitrary charge network. For example, if the initial ket is such that at least one of the spacings $\Delta^{+}=\alpha n^+$ at some vertex $v$ is labelled by the minimum integer 1,  i.e. $\Delta=\alpha$, and the boost parameter $e^{\lambda}$ is any rational number such that $e^{\lambda}<1$, the boosted state as obtained from the initial one does not lie in the kinematic Hilbert space. Thus, global boosts, unless state-dependent, cannot be implemented in the quantum theory\footnote{It would be worth exploring the implications of possible state dependent isometries in the quantum theory. We do not attempt such an analysis here.}. 

To conclude, only continuous global translations, which form a subgroup of the global Poincare group, are realized in the quantum theory. However, such a violation of the symmetries would be blurred as long as one is concerned with states whose $k$ spacings are much larger than the minimum integer, and stays so even after the application of boosts. This implies that Lorentz invariance can still be maintained, but only in an effective sense.
\subsection{Group averaging} 
Here, we will adopt a simpler notation by suppresing the `$\pm$' indices.  For any state $\psi_{s_e,\m{X},s_m}=\ket{s_e,\m{X}}\otimes \ket{s_m}$ belonging to the kinematic Hilbert space $H_{kin}$, a group averaging map can be used construct a physical Hilbert space\cite{almt}. Let us consider all the distinct gauge images $\psi_{s_e,\m{X},s_m}^{\xi}=\ket{s_e,\m{X}}_{\xi}\otimes \ket{s_m}_{\xi}$ obtained through the action of $\hat{U}_{\xi}$ on $\psi_{s_e,\m{X},s_m}$. Then, a formal solution of the constraints is given by the sum over all the distinct gauge images: 
\begin{eqnarray}\label{etamap}
 \bra{\eta[\psi_{s_e,\m{X},s_m}]}=\eta_{[s_e,\m{X},s_m]}\sum_{\xi \in G[s_e,\m{X},s_m]}~_{\xi}\bra{s_m}~\otimes~_{\xi}\bra{s_e,\m{X}} 
\end{eqnarray}
where, the positive real coefficient $\eta_{[s_e,\m{X},s_m]}$ depends only on the gauge orbit of $\ket{s_e,\m{X}}\otimes\ket{s_m}$. The physical state $\bra{\eta[\psi_{s_e,\m{X},s_m}]}$ lies in a larger space $\Phi^{*}\supset H_{kin}$.
The corresponding inner product between the physical states is defined as:
\begin{eqnarray*}
 \braket{\eta[\psi_{s_e,\m{X},s_m}]|\eta[\psi_{s'_e,\m{X'},s'_m}]}_{phy}~=~\bra{\eta[\psi_{s_e,\m{X},s_m}]}[\psi_{s'_e,\m{X'},s'_m}]
\end{eqnarray*}

In order to reduce the ambiguity in the coefficients $\eta_{[s_e,\m{X},s_m]}$, one can use the commutativity of the $\eta$ map with any Dirac observable in the theory. However, the analysis is very similar to the case of PFT on a compact spatial slice as studied in \cite{sengupta}, and does not lead to a significant reduction of the ambiguity. Since this exercise does not provide any illuminating detail in our case, we choose to omit a discussion of that. As in the compact case, the physical space is nonseparable, since the gauge invariant data corresponding to each physical basis state contains the real numbers $\m{X}(v)$ at the vertices $v$ (for a detailed discussion in this regard, see \cite{sengupta}), which are uncountable labels.
\section{Quantum geometry of spacetime}
A physical state as obtained through the averaging can be interpreted as a quantum spacetime, whose coordinates are $X^{\pm}$. Let us assume that such a state is a sum of all distinct gauge images of any charge network state $\ket{\gamma_e^{\pm},\overrightarrow{k},\m{X}^{\pm}}\otimes \ket{\gamma_m^{\pm},\overrightarrow{l}}$. For this state, the total embedding charge $k^{\pm}_{n,tot}=k_n^{\pm}+\m{X}^{\pm}$ for the $n$-th edge spans the continuous range [$k_n^{\pm} +\m{X}^{\pm}_{min},~k_n^{\pm} +\m{X}^{\pm}_{max}$], where $\m{X}^{\pm}_{min}$ and $\m{X}^{\pm}_{max}$ denote the minimum and maximum values of $\m{X}^{\pm}$ along the edge, respectively. However, each such continuum is followed by a jump in the value of $k^{\pm}_{n,tot}$. Now, note that the continuous interval [$k_n^{\pm} +\m{X}^{\pm}_{min},~k_n^{\pm} +\m{X}^{\pm}_{max}$] represents a spacetime strip, finite along the `$\pm$' direction and infinite along the `$\mp$' direction. Thus, intersection of any pair of intervals [$k_n^{+}+\m{X}^{+}_{min},~k_n^{+}+\m{X}^{+}_{max}]$ and [$k_n^{-}+\m{X}^{-}_{min},~k_n^{-}+\m{X}^{-}_{max}]$ corresponds to a rectangular strip of flat spacetime. The resulting physical spacetime is composed of discrete rectangular strips, followed by voids. The matter charges sit as discrete points on these strips. However, such a discrete structure prevails only at the interior, which spans a compact region. As one approaches asymptotia, one obtains a physical state characterised by a smoothly varying embedding charge. Thus, the asymptotic quantum geometry resembles a classical continuum. 

\section{Concluding remarks}
We have constructed a canonical quantization of the theory of a parametrized scalar field on noncompact spatial slices, with a view to analyse the quantization of asymptotically flat gravity. The methods adopted here are based on loop (polymer) quantization techniques. The most important ingredient in this construction is a new kinematical representation which corresponds to a quantum geometry with a nondegenerate vacuum metric. The representation of this kind was originally introduced in the context of Loop Quantum Gravity\cite{k,s}, in order to incorporate the notion of a smooth effective spacetime within the quantum kinematics. In the case of PFT, the states in this representation have continuous embedding labels in addition to the discrete embedding and matter labels. Using the smooth background structure underlying this construction, the boundary conditions for the basic fields at spatial infinity were shown to be consistently implemented in the quantum theory. This signifies an important step, since it is not known how to impose the asymptotic conditions within the standard representation with a degenerate vacuum geometry, which is devoid of any smooth structure. 

The action of PFT is invariant under the group of conformal isometries, which generate Weyl rescalings of the metric. This contains the global Poincare group consisting of translations and boosts. However, in the quantum theory, only the continuous global translations, which characterise a subgroup of the Poincare group, are realized. This happens to be exactly the same subgroup as is obtained in the quantization of two dimensional PFT on a compact spatial slice, as analyzed in \cite{sengupta}. However, it is still possible to have state dependent boosts within this formulation. If such a nontrivial `symmetry group' of conformal isometries is treated to be more fundamental than the classical one, it is to be seen how the quantum theory responds to such a viewpoint. This issue requires a deeper study. 

The quantum spacetime, as determined by a physical state in the quantum theory, is made up of discrete rectangular strips at the interior region, and exhibits a smooth geometry at the asymptotia. We emphasize that such a feature cannot be realized within the standard loop (polymer) representation with a degenerate vacuum geometry. It is true that one still needs to understand the details regarding the possible coarse-graining which leads to the effective smoothness of the charge-network states at the spatial infinity. In other words, one must ask whether there is a way to implement the boundary conditions and recover the spacetime continuum at asymptotia directly within the discrete setting without using any extra structure such as a smooth embedding. However, we believe that even if a complete understanding of such issues is missing at this stage, it might be worthwhile to generalize the framework as set up here to the case of four dimensional gravity theory on noncompact spatial slices. 
%
%
%

\acknowledgments
I must thank M.Varadarajan for suggesting the line of research as carried out in this paper, and for his critical comments and encouragement. I am indebted to M.Campiglia and A.Laddha for useful discussions as well as their careful reading of the manuscript. Comments of M. Bojowald and H.Sahlmann on this work are also gratefully acknowledged.

%

\end{document}